\magnification=\magstep0
\font\bbf=cmbx10 scaled\magstep1
\hsize=6.45truein

\topglue 2in
\centerline{\bbf COMPUTATION OF TWO-BODY MATRIX ELEMENTS}
\centerline{\bbf FROM THE ARGONNE $v_{18}$ POTENTIAL
  \footnote*{{\it work supported by DOE grant DE-FG02-87ER-40371}}}
\smallskip  
\centerline{by}
\smallskip  
\centerline{Bogdan~Mihaila and Jochen~H.~Heisenberg}
\centerline{\it Dept. of Physics, University of New Hampshire, 
Durham, N.H. 03824}
\vskip 0.5in

\noindent
{\it Abstract.} We discuss the computation of two-body matrix elements
from the Argonne $v_{18}$ interaction.  
The matrix elements calculation is presented both 
in particle-particle and in particle-hole angular momentum coupling. 
The procedures developed here can be applied to the case of other 
{\it NN} potentials, provided that they have a similar operator format.
\vfill

\beginsection{0. Introduction}

The Argonne $v_{18}$ potential~[1] is an updated version of the nonrelativistic
Argonne potential that fits both {\it np} data and {\it pp} data,
as well as low-energy {\it nn} data scattering parameters and 
deuteron properties.
The potential was directly fit to the Nijmegen {\it NN} scattering database, 
which contains 1787 {\it pp} and 2514 {\it np} data in the range 0-350 MeV, 
and has an excellent $\chi^2$ per datum of 1.09 .

The strong interaction part of the potential is projected into 
an operator format with 18 terms:
A charge-independent part that has 14 operator components 
(as in the older Argonne $v_{14}$ )
$$ 1, \ \ \sigma_i\cdot\sigma_j, \ \ S_{ij}, \ L\cdot S, \ \ 
   L^2, \ \ L^2 \ \sigma_i\cdot\sigma_j, \ \ (L\cdot S)^2 $$
$$ \tau_i\cdot\tau_j, \ \ (\sigma_i\cdot\sigma_j) \ (\tau_i\cdot\tau_j), \ \
   S_{ij} \ (\tau_i\cdot\tau_j), \ L\cdot S \ (\tau_i\cdot\tau_j), \ \ 
   L^2 (\tau_i\cdot\tau_j), \ \ 
   L^2 \ (\sigma_i\cdot\sigma_j) \ (\tau_i\cdot\tau_j), \ \ 
   (L\cdot S)^2 \ (\tau_i\cdot\tau_j) $$
And a charge-independence breaking part that has 
three charge-dependent operators
$$ T_{ij}, \ \ (\sigma_i\cdot\sigma_j) \ T_{ij}, \ \ S_{ij} \ T_{ij} $$
where $T_{ij} = 3 \tau_{zi} \tau_{zj} - \tau_i \cdot \tau_j$ is 
the isotensor operator, defined analogous to the $S_{ij}$ operator; 
and one charge-asymmetric operator
$$ \tau_{zi} + \tau_{zj} $$

The potential includes also a complete electromagnetic potential, 
containing Coulomb, Darwin-Foldy, vacuum polarization, and magnetic moment terms
with finite-size effects.

\vskip 0.2in
Matrix elements of the {\it NN} interaction can be specified either 
in particle-particle ({\it pp}) coupling or 
in particle-hole ({\it ph}) coupling. 
Both of the matrix elements completely specify the interaction, 
and either set can be calculated from the other set. 
We first discuss the symmetries and the angular momentum couplings (AMC).
\vskip 0.2in

\beginsection{1. Symmetries of two-body matrix elements}

Generally we use the definition: $k=j+1/2$ and $\hat j=\sqrt{2j+1}$. 
The matrix element in m-representation can be written as
$$\langle m_a~m_b~\vert V\vert~m_c~m_d\rangle~=V_{m_am_b,m_cm_d}
 =\langle m_am_c^{-1}\vert V\vert m_dm_b^{-1}\rangle=
  \langle m_a\bar m_c\vert V\vert m_d\bar m_b\rangle\eqno{1.1}$$
We use these notations interchangeably. 

The angular momentum coupled matrix element is constructed as
$$\langle(j_aj_b)_L~\vert V^L\vert~(j_cj_d)_L\rangle=\sum_{m_am_bm_cm_d}
\langle j_a m_a j_b m_b\vert LM\rangle\langle j_c m_c j_d m_d\vert
LM\rangle \langle m_am_b\vert V \vert m_cm_d\rangle\eqno{1.2}$$
which is independent of $M$. It has the symmetry:
$$\langle(j_aj_b)_L~\vert V^L\vert~(j_cj_d)_L\rangle=
(-)^{k_a+k_b+k_c+k_d}\langle(j_bj_a)_L~\vert V^L\vert~(j_dj_c)_L\rangle$$
The inverse relation is
$$\langle m_am_b\vert V \vert m_cm_d\rangle=\sum_{LM}
\langle j_a m_a j_b m_b\vert LM\rangle\langle j_c m_c j_d m_d\vert
LM\rangle \langle (j_aj_b)_L\vert V^L \vert (j_cj_d)_L\rangle\eqno{1.3}$$

Instead of using this {\it pp}-coupling one can use {\it ph}-coupling 
to construct angular momentum coupled matrix elements.
The {\it ph}-coupled matrix element is defined as
$$\langle(j_1\bar j_2)_{\lambda}\vert V^{\lambda}\vert(j_3\bar j_4)_
{\lambda}\rangle =\sum_{m_1m_2m_3m_4}(-)^{k_2+m_2+k_4-m_4}
\langle j_1m_1j_2-m_2\vert\lambda\mu\rangle
\langle j_3m_3j_4-m_4\vert\lambda\mu\rangle
\langle m_1m_2^{-1}\vert V\vert m_3m_4^{-1}\rangle\eqno{1.4}$$
These are again independent of $\mu$.

The matrix elements in {\it pp}-coupling are related to the matrix
elements in {\it ph}-coupling through~[2]
$$\langle (j_1\bar j_2)_{\lambda}\vert V^{\lambda}\vert (j_3\bar j_4)_{\lambda}
\rangle=\sum_L(-)^{k_2+k_3+L+1}(2L+1)\Bigl\lbrace\matrix{j_1&j_2&\lambda\cr
j_3&j_4&L}\Bigr\rbrace
\langle (j_1j_4)_L\vert V^L\vert (j_2j_3)_L\rangle\eqno{1.5}$$
This is obtained via general recoupling, using:
$$\eqalign{
\sum_M&\langle j_1m_1j_2m_2|LM\rangle\langle j_3m_3j_4m_4|LM\rangle=\cr
=&(-)^{j_2+m_1+j_4+m_3}\sum_{\lambda \mu}(2L+1)
\Bigl\lbrace\matrix{j_1&j_2&L\cr j_3&j_4&\lambda}\Bigr\rbrace
\langle j_1m_1j_4-m_4|\lambda\mu\rangle\langle j_3m_3j_2-m_2|
\lambda\mu\rangle \cr
=&(-)^{j_2-m_2+j_4+m_3+L}\sum_{\lambda \mu}(2L+1)
\Bigl\lbrace\matrix{j_1&j_2&L\cr j_4&j_3&\lambda}\Bigr\rbrace
\langle j_1m_1j_3-m_3|\lambda\mu\rangle\langle j_4m_4j_2-m_2|
\lambda\mu\rangle }$$
The inverse relation for the matrix elements is
$$\langle (j_1j_4)_L\vert V^{L}\vert
(j_2j_3)_L\rangle=(-)^{L+1}\sum_{\lambda} (-)^{k_2+k_3} (2\lambda+1)
\Bigl\lbrace\matrix{j_1&j_2&\lambda\cr j_3&j_4&L}\Bigr\rbrace
\langle(j_1\bar j_2)_{\lambda}\vert V^{\lambda}\vert (j_3\bar
j_4)_{\lambda}\rangle\eqno{1.6}$$

Anti-symmetric matrix elements are formed in m-representation as
$$\langle m_am_b\vert V^a\vert m_cm_d\rangle=
  \langle m_am_b\vert V  \vert m_cm_d\rangle-
  \langle m_am_b\vert V  \vert m_dm_c\rangle\eqno{1.7}$$
In {\it pp} angular momentum coupling this becomes
$$\langle(j_aj_b)_L\vert V^{aL}\vert (j_cj_d)_L\rangle=
  \langle(j_aj_b)_L\vert V^L\vert (j_cj_d)_L\rangle+(-)^{L+k_c+k_d}
  \langle(j_aj_b)_L\vert V^L\vert (j_dj_c)_L\rangle\eqno{1.8}$$
These anti-symmetric matrix elements have the symmetry
$$\langle(j_aj_b)_L\vert V^{aL}\vert (j_cj_d)_L\rangle=
(-)^{L+k_c+k_d}  \langle(j_aj_b)_L\vert V^{aL}\vert (j_dj_c)_L\rangle=
(-)^{L+k_a+k_b}  \langle(j_bj_a)_L\vert V^{aL}\vert (j_cj_d)_L\rangle
\eqno{1.9}$$
In {\it ph}-coupling the anti-symmetrization can be obtained by
transforming the anti-symmetric matrix elements in {\it pp}-coupling
using (1.5). Thus we obtain
$$\eqalign{\langle (j_1\bar j_2)_{\lambda}\vert V^{a\lambda}\vert (j_3\bar j_4)
_{\lambda} \rangle&=
  \langle (j_1\bar j_2)_{\lambda}\vert V^{\lambda}\vert (j_3\bar j_4)
 _{\lambda} \rangle+ \cr &\sum_{\lambda'}(-)^{\lambda+\lambda'}
(-)^{k_2+k_3} (2\lambda'+1)
\Bigl\lbrace\matrix{j_1&j_2&\lambda\cr j_4&j_3&\lambda'}\Bigr\rbrace
  \langle (j_1\bar j_3)_{\lambda'}\vert V^{\lambda'}\vert (j_2\bar j_4)
 _{\lambda'} \rangle}\eqno{1.10}$$

The symmetry of the relations can be significantly improved by
introducing the {\it Ring}-phase. 
For {\it pp}-coupling the {\it Ring}-phase is
$$\langle (j_aj_b)_L\vert V^{aLR}\vert (j_cj_d)_L\rangle=(-)^{k_a+k_c}
  \langle (j_aj_b)_L\vert V^{aL}\vert (j_cj_d)_L\rangle\eqno{1.11}$$
With this extra phase the matrix elements have the symmetry
$$\langle (j_aj_b)_L\vert V^{aLR}\vert (j_cj_d)_L\rangle=
  (-)^L \langle (j_aj_b)_L\vert V^{aLR}\vert (j_dj_c)_L\rangle=
  (-)^L \langle (j_bj_a)_L\vert V^{aLR}\vert (j_cj_d)_L\rangle\eqno{1.12}$$
and the anti-symmetric matrix elements can be calculated as
$$\langle(j_aj_b)_L\vert V^{aLR}\vert (j_cj_d)_L\rangle=
  \langle(j_aj_b)_L\vert V^{LR}\vert (j_cj_d)_L\rangle+(-)^L
  \langle(j_aj_b)_L\vert V^{LR}\vert (j_dj_c)_L\rangle\eqno{1.13}$$

For matrix elements in {\it ph}-coupling the {\it Ring}-phase is
$$\langle (j_1\bar j_2)_{\lambda}\vert V^{a\lambda R}\vert(j_3\bar j_4)_
{\lambda}\rangle=(-)^{k_1+k_3}
  \langle (j_1\bar j_2)_{\lambda}\vert V^{a\lambda}\vert(j_3\bar j_4)_
{\lambda}\rangle\eqno{1.14}$$
With this phase the anti-symmetric matrix elements 
in {\it ph}-coupling have the symmetry
$$\eqalign{
\langle (j_1\bar j_2)_{\lambda}\vert V^{a\lambda R}\vert(j_3\bar j_4)_
{\lambda}\rangle&=
  \langle (j_2\bar j_1)_{\lambda}\vert V^{a\lambda R}\vert(j_4\bar j_3)_
{\lambda}\rangle\cr
&=\sum_{\lambda'} (-)^{\lambda+\lambda'}(2\lambda'+1)\Bigl\lbrace
\matrix{j_1&j_3&\lambda'\cr j_4&j_2&\lambda}\Bigr\rbrace
\langle (j_1\bar j_3)_{\lambda'}\vert V^{a\lambda' R}\vert(j_2\bar j_4)_
{\lambda'}\rangle}\eqno{1.14}$$
Also, the anti-symmetric matrix elements in {\it ph}-coupling can be 
constructed as
$$\eqalign{\langle (j_1\bar j_2)_{\lambda}\vert V^{a\lambda R}\vert 
(j_3\bar j_4) _{\lambda} \rangle&=
  \langle (j_1\bar j_2)_{\lambda}\vert V^{\lambda R}\vert (j_3\bar j_4)
 _{\lambda} \rangle+ \cr &\sum_{\lambda'}(-)^{\lambda+\lambda'}
 (2\lambda'+1)
\Bigl\lbrace\matrix{j_1&j_2&\lambda\cr j_4&j_3&\lambda'}\Bigr\rbrace
  \langle (j_1\bar j_3)_{\lambda'}\vert V^{\lambda'R}\vert (j_2\bar j_4)
 _{\lambda'} \rangle}\eqno{1.15}$$
and the relation between {\it pp}-coupling and {\it ph}-coupling
is simplified to
$$\langle (j_1j_4)_L\vert V^{aLR}\vert(j_2j_3)_L\rangle= (-)^{L+1}\sum_
{\lambda} (2\lambda+1)\Bigl\lbrace\matrix{j_1&j_2&\lambda\cr j_3&j_4&L}
\Bigr\rbrace
\langle(j_1\bar j_2)_{\lambda}\vert V^{a\lambda R}\vert(j_3\bar j_4)_
{\lambda}\rangle \eqno{1.16}$$
or its reverse relation
$$\langle (j_1\bar j_2)_{\lambda}\vert V^{a\lambda R}\vert 
(j_3\bar j_4)_{\lambda}
\rangle=\sum_L(-)^{L+1}(2L+1)\Bigl\lbrace\matrix{j_1&j_2&\lambda\cr
j_3&j_4&L}\Bigr\rbrace
\langle (j_1j_4)_L\vert V^{aLR}\vert (j_2j_3)_L\rangle\eqno{1.17}$$

In all the nuclear structure calculations we will use anti-symmetric 
matrix elements that
include the {\it Ring}-phase. For simplicity we will drop the
superscripts $aR$ throughout.
\vskip 0.2in

\beginsection{2. General considerations and relationships}

We assume the interaction can be written 
as a scalar product of tensors of rank $(k)$. 
Then the ph-ph matrix element  including the
{\it Ring}-phase of $(-)^{(k_1+k_3)}$  can be computed as
$$\eqalign{
<(1,\bar 2)_{\lambda}|V^R|(3,\bar 4)_{\lambda}>=&
(-)^{(k_1+k_3)}<(1,\bar 2)_{\lambda}|V|(3,\bar 4)_{\lambda}>=\cr
=&(-)^{(k_1+k_3)}<(1,\bar 2)_{\lambda}|\bigl[U^{(k)}(1)\odot
V^{(k)}(2)\bigr]|(3,\bar 4)_{\lambda}>=\cr
=&\delta_{k,\lambda}
 (-)^{k_1}{1\over{\hat \lambda}}<1\Vert U^{(\lambda)}\Vert 2>~~
  (-)^{k_4}{1\over{\hat \lambda}}<4\Vert V^{(\lambda)}\Vert 3>}\eqno{2.1}$$
Thus it is necessary to bring the various interactions into this form.

We will use the Fourier transform to separate the variables $\vec r_1$ and
$\vec r_2$. For this, the general relation can be worked out :
$$\eqalign{V(r_{12})Y_{J,M}(\hat r_{12})=&
{2\over{\pi}}\int q^2dq \tilde V^J(q)\sum_{\ell_1,\ell_2}
{{\hat \ell_1\hat\ell_2}\over{\hat J}}{1\over{\sqrt{4\pi}}}
\langle \ell_10\ell_20\vert J0\rangle\cr
\times &j_{\ell_1}(qr_1)j_{\ell_2}(qr_2)
(i)^{(\ell_1-\ell_2-J)}4\pi \bigl[ Y_{\ell_1}(\hat r_1)\otimes
Y_{\ell_2}(\hat r_2)\bigr]^{(J,M)}}\eqno{2.2}$$
where the form factor of the interaction is
$$\tilde V^J(q)=\int V(r_{12})j_J(qr_{12})r_{12}^2dr_{12}\eqno{2.3}$$

The various necessary operators are defined as
$$\eqalign{\vec r_{12}&=\vec r_1-\vec r_2\cr
\vec S&={1\over{2}}(\vec\sigma_1+\vec\sigma_2)\cr
\vec L&={1\over(2i)}\vec r_{12}\times (\vec\nabla_1-\vec\nabla_2)
}\eqno{2.4}$$
Using ([3], eq. 5.1.4)
$$\hat r_{12}^{(1)}=\sqrt{{{4\pi}\over{3}}}Y^{(1)}(\hat r_{12})\eqno{2.5}$$
and ([3] eq. 5.2.4)
$$a^{(k)}\odot b^{(k)}=(-)^k \hat k \bigl[ a^{(k)}\otimes b^{(k)}\bigr]^
{(0)}\eqno{2.6}$$
Using ([3] eq. 5.1.8) we write
$$\eqalign{
L^{(1)}=
      &{1\over{\sqrt{2}}}r_{12}\bigl[\hat r_{12}^{(1)}\otimes
        (\vec\nabla_1-\vec\nabla_2)^{(1)}\bigr]^{(1)}\cr
=&r_{12}\sqrt{{{2\pi}\over{3}}}\bigl[Y^{(1)}(\hat
r_{12})\otimes(\vec\nabla_1-\vec\nabla_2)^{(1)}\bigr]^{(1)}
}\eqno{2.7}$$

For terms containing $L^2$ we need to evaluate the commutator
for the tensor components
with $\vec\nabla=\vec\nabla_1-\vec\nabla_2$
which takes the form
$$\bigl[ \vec r^{(1)}_s, \vec\nabla^{(1)}_{-t} \bigr]=2\delta_{s,t}$$
With this we obtain
$$\eqalign{\Bigl[\bigl[\vec r \otimes \vec\nabla \bigr]^{(1)} \otimes \bigl[
  \vec r \otimes \vec\nabla \bigr]^{(1)}\Bigr]^{(j)}=&
  3\sum_{J_r,J_p}\hat J_r \hat J_p
  \Biggl\lbrace\matrix{1&1&1\cr 1&1&1\cr J_r&J_p&j}\Biggr\rbrace
  \Bigl[\bigl[ \vec r \otimes \vec r\bigr]^{(J_r)}\otimes
  \bigl[ \vec \nabla \otimes \vec \nabla\bigr]^{(J_p)}\Bigr]^{(j)}\cr
  &+6(-)^{(j+1)}
  \Bigl\lbrace\matrix{1&1&1\cr 1&1&j}\Bigr\rbrace
  \bigl[ \vec r \otimes \vec \nabla\bigr]^{(j)}
  }\eqno{2.8}$$
Due to vanishing $\vec a \times \vec a$, each of $j,J_r,J_p$ can take
up only the values of 0 or 2.
For this operator the following relation is useful
$$\bigl[\vec r \otimes \vec r\bigr]^{(J)}=
r^2{{4\pi}\over{3}}\bigl[ Y^{(1)}(\hat r)\otimes Y^{(1)}(\hat r)\bigr]
^{(J)}=r^2\sqrt{{{4\pi}\over{2J+1}}}<1010|J0>Y^{(J)}(\hat
r)\eqno{2.9}$$
Thus we can write
$$\eqalign{
\Bigl[\bigl[\vec r \otimes \vec\nabla \bigr]^{(1)} \otimes \bigl[
  \vec r \otimes \vec\nabla \bigr]^{(1)}\Bigr]^{(j)}=&
  3r^2\sum_{J_r,J_p} \hat J_p
  \Biggl\lbrace\matrix{1&1&1\cr 1&1&1\cr J_r&J_p&j}\Biggr\rbrace
  \langle 1~0~1~0\vert J_r~0\rangle
  \sqrt{4\pi}\Bigl[Y^{(J_r)}(\hat r)\otimes
  \bigl[ \vec \nabla \otimes \vec \nabla\bigr]^{(J_p)}\Bigr]^{(j)}\cr
  &-6r \Bigl\lbrace\matrix{1&1&1\cr 1&1&j}\Bigr\rbrace 
  \sqrt{{{4\pi}\over{3}}}
  \bigl[ Y^{(1)}(\hat r) \otimes \vec \nabla\bigr]^{(j)}
  }\eqno{2.10}$$

\beginsection{3. Central interaction}

Matrix elements for the interaction term
$V^c(r_{12}).$ This term includes the Coulomb interaction.
The matrix element contributes only for natural parity matrix elements,
i.e. $\lambda+l_1+l_2=even$. Including the ``Ring''-phase of
$(-)^{k_1+k_3}$ we get
$$\langle (1\bar 2)_{\lambda}\vert V^{c,R} \vert (3\bar 4)_{\lambda}\rangle=
{2\over{\pi}} \int q^2dq~F^c(q)~
\Biggl({\sqrt{4\pi}\over{\hat\lambda}}(-)^{k_1}
\langle 1\Vert Y_{\lambda}\Vert 2\rangle
      {\sqrt{4\pi}\over{\hat\lambda}}(-)^{k_4}
      \langle 4\Vert Y_{\lambda}\Vert 3\rangle\Biggr)
 \eqno{3.1}$$
where the form factor for the central interaction is
$$F^c(q)=\int V^c(r_{12})~j_0(qr_{12})~r_{12}^2dr_{12}\eqno{3.2}$$

For the evaluation of the matrix element we need the following
reduced matrix elements in which we include the radial matrix elements
involving the Bessel functions: These also contain the {\it Ring}-phase.
They are evaluated for terms that do not contain $\sigma$ using ([3] eq. 7.1.7)
$${\sqrt{4\pi}\over{\hat k}}(-)^{k_1}<(\ell_1s_1)j_1\Vert G^{(k)}\Vert
(\ell_2s_2)j_2>=(-)^{(\ell_1+k_1+k_2)}
{{\hat j_1\hat j_2}\over{\hat k}}\Bigl
\lbrace\matrix{\ell_1&\ell_2&k\cr j_2&j_1&1/2}\Bigr\rbrace
\sqrt{4\pi}<\ell_1\Vert G^{(k)}\Vert \ell_2>\eqno{3.3}$$
We abbreviate this as
$$f(k,1,2)\sqrt{4\pi}<\ell_1\Vert G^{(k)}\Vert \ell_2>\eqno{3.4}$$
At this point we need the following reduced matrix elements ([3] eq. 5.4.5):
$$\sqrt{4\pi}<\ell_1\Vert Y^{\ell}\Vert \ell_2>=
(-)^{\ell_1}\hat \ell_1\hat \ell_2
<\ell_10\ell_20\vert\ell 0>
\langle R_1(r)\vert j_{\lambda}(qr)\vert R_2(r)\rangle\eqno{3.5}$$
We abbreviate this as 
$$y(\ell,1,2)\langle R_1(r)\vert j_{\lambda}(qr)\vert R_2(r)\rangle
\eqno{3.6}$$
With these
definitions we write the reduced matrix elements as
$${\sqrt{4\pi}\over{\hat\lambda}}(-)^{k_1}
  \langle 1\Vert Y^{\lambda} \Vert 2\rangle=
  f(\lambda,1,2)y(\lambda,1,2)
 \langle R_1(r)\vert j_{\lambda}(qr)\vert R_2(r)\rangle\eqno{3.7}$$

\beginsection{4. Term (LL)}

We write the interaction using eq.(2.8) with $j=0$ as
$$\eqalign{
V=V^{L2}&(r_{12})\bigl[ L\odot L\bigr]=-\sqrt{3}V^{L2}(r_{12})
   \bigl[ L^{(1)}\otimes L^{(1)}\bigr]^{(0)}\cr
   =&-{\sqrt{3}\over{2}}V^{L2}(r_{12})\Bigl[\bigl[ \vec r_{12}\otimes
   \vec\nabla_{12}\bigr]^{(1)}\otimes\bigl[\vec r_{12}\otimes\vec\nabla
   _{12}\bigr]^{(1)}\Bigr]^{(0)}\cr
   =&-{\sqrt{3}\over{2}}r_{12}^2V^{L2}(r_{12})\sum_{J_r,J_p}
   \langle 1010|J_r0\rangle \sqrt{4\pi}~3\hat J_p
   \Biggl\lbrace\matrix{1&1&1\cr 1&1&1\cr J_r&J_p&0}\Biggr\rbrace
   \Bigl[ Y^{(J_r)}(\hat r_{12})\otimes\bigl[\vec\nabla_{12}\otimes
   \vec\nabla_{12}\bigr]^{(J_p)}\Bigr]^{(0)}\cr
   &~~~+{6\over{2}}\sqrt{4\pi}~
   \Bigl\lbrace\matrix{1&1&1\cr 1&1&0}\Bigr\rbrace
   r_{12}V^{L2}(r_{12})\bigl[Y^{(1)}(\hat r_{12})\otimes
   \vec\nabla_{12}\bigr]^{(0)}\cr
   =&{3\over{2}} \sqrt{4\pi}~ r_{12}^2V^{L2}(r_{12})\sum_{J}
   \langle 1010|J0\rangle
   \Bigl\lbrace\matrix{1&1&1\cr 1&1&J}\Bigr\rbrace
   \Bigl[ Y^{(J)}(\hat r_{12})\otimes\bigl[\vec\nabla_{12}\otimes
   \vec\nabla_{12}\bigr]^{(J)}\Bigr]^{(0)}\cr
   &~~~-\sqrt{4\pi}~
   r_{12}V^{L2}(r_{12})\bigl[Y^{(1)}(\hat r_{12})\otimes
   \vec\nabla_{12}\bigr]^{(0)}
   }\eqno{4.1}$$
As the $9j$-coefficient requires $J_r=J_p$ we have set
$J_r=J_p=J$. $J$ can assume the values of $0$ or $2$.
We employ the Fourier transform where we define the form factors of
this interaction as
$$\tilde V^{L2,J}(q)=\int r_{12}^4V^{L2}(r_{12})j_J(qr_{12})dr_{12}
\eqno{4.2}$$
and similarly
$$\bar V^{L2,1}(q)=\int r_{12}^3V^{L2}(r_{12})j_1(qr_{12})dr_{12}
\eqno{4.3}$$
Then, using eq.(2.2) with eq.(4.1) we obtain
$$\eqalign{
V=&\sum_J{3\over{2}}\langle 1010|J0\rangle {1\over{\hat J}}
  \Bigl\lbrace\matrix{1&1&1\cr 1&1&J}\Bigr\rbrace\sum_{\ell_1,\ell_2}
  (-)^{(\ell_1+\ell_2+J)/2}\hat \ell_1\hat \ell_2
  <\ell_10\ell_20|J0>
  {2\over{\pi}}\int q^2dq \tilde V^{L2J}(q)
  j_{\ell_1}(qr_1)j_{\ell_2}(qr_2)\cr
  ~&~~~~~~~~\times(-)^{\ell_2}4\pi\Bigl[\bigl[Y^{(\ell_1)}(\hat
  r_1)\otimes Y^{(\ell_2)}(\hat r_2)\bigr]^{(J)}
  \otimes\bigl[\vec\nabla_{12}\otimes
  \vec\nabla_{12}\bigr]^{(J)}\Bigr]^{(0)}\cr
  ~&+  \sum_{\ell_1,\ell_2}
  (-)^{(\ell_1+\ell_2+1)/2}{{\hat \ell_1\hat \ell_2}\over{\sqrt{3}}}
  <\ell_10\ell_20|10> {2\over{\pi}} \int q^2dq
  \bar V^{L2}(q) j_{\ell_1}(qr_1)j_{\ell_2}(qr_2) \cr
  ~&~~~~~~~~\times(-)^{\ell_2}4\pi\Bigl[\bigl[Y^{(\ell_1)}(\hat
  r_1)\otimes Y^{(\ell_2)}(\hat r_2)\bigr]^{(1)}
  \otimes\vec\nabla_{12}\Bigr]^{(0)}
  }\eqno{4.4}$$
Substituting $\vec\nabla_{12}=\vec\nabla_1-\vec\nabla_2$ results in
five terms which need to be recoupled in order to be of the form
$F^{(k)}(1)\odot G^{(k)}(2)$. These are
$$\eqalign{
Z_1=&(-)^{\ell_2}
   \Bigl[\bigl[Y^{(\ell_1)}(1)\otimes Y^{(\ell_2)}(2)\bigr]^{(J)}\otimes
          \bigl[\vec\nabla_1 \otimes
          \vec\nabla_1\bigr]^{(J)}\Bigr]^{(0)}\cr
   =&{1\over{\hat \ell_2}}\biggl[
   \Bigl[Y^{(\ell_1)}(1)\otimes \bigl[\vec\nabla_1\otimes
   \vec\nabla_1\bigr]^{(J)}\Bigr]^{(\ell_2)}\odot Y^{(\ell_2)}(2) \biggr]\cr
Z_2=&(-2)(-)^{\ell_2}
   \Bigl[\bigl[Y^{(\ell_1)}(1)\otimes Y^{(\ell_2)}(2)\bigr]^{(J)}\otimes
          \bigl[\vec\nabla_1 \otimes
          \vec\nabla_2\bigr]^{(J)}\Bigr]^{(0)}\cr
   =&2\sum_k\hat J
   \Bigl\lbrace\matrix{\ell_1&\ell_2&J\cr 1&1&k}\Bigr\rbrace
   \Bigl[\bigl[Y^{(\ell_1)}(1)\otimes\vec\nabla_1\bigr]^{(k)}\odot
         \bigl[Y^{(\ell_2)}(2)\otimes\vec\nabla_2\bigr]^{(k)}\Bigr]\cr
}$$
$$\eqalign{
Z_3=&(-)^{\ell_2}
  \Bigl[\bigl[Y^{(\ell_1)}(1)\otimes Y^{(\ell_2)}(2)\bigr]^{(J)}\otimes
          \bigl[\vec\nabla_2 \otimes
          \vec\nabla_2\bigr]^{(J)}\Bigr]^{(0)}\cr
             =&{1\over{\hat \ell_1}}\biggl[Y^{(\ell_1)}(1)\odot
   \Bigl[Y^{(\ell_2)}(2)\otimes \bigl[\vec\nabla_2\otimes
   \vec\nabla_2\bigr]^{(J)}\Bigr]^{(\ell_1)}\biggr]\cr
Z_4=&(-)^{\ell_2} \Bigl[\bigl[Y^{(\ell_1)}(1)\otimes
Y^{(\ell_2)}(2)\bigr]^{(1)}\otimes \vec\nabla_1\Bigr]^{(0)}\cr
   =&{1\over{\hat \ell_2}}\Bigl[
   \bigl[Y^{(\ell_1)}(1)\otimes\vec\nabla_1\bigr]^{(\ell_2)}\odot
   Y^{(\ell_2)}(2)\Bigr] \cr
Z_5=&-(-)^{\ell_2} \Bigl[\bigl[Y^{(\ell_1)}(1)\otimes
Y^{(\ell_2)}(2)\bigr]^{(1)}\otimes \vec\nabla_2\Bigr]^{(0)}\cr
   =&{1\over{\hat \ell_1}}\Bigl[Y^{(\ell_1)}(1)\odot
   \bigl[Y^{(\ell_2)}(2)\otimes\vec\nabla_2\bigr]^{(\ell_1)}\Bigr]
}\eqno{4.5}$$

For the evaluation of this matrix element we need the additional
reduced matrix elements:
The matrix element involving the operator $\nabla$: 
$$\eqalign{\sqrt{4\pi}
<\ell_1\Vert (Y^{\ell}\nabla)^{(\kappa)}\Vert \ell_2>=&
(-)^{\ell_2+\kappa}\hat \ell_1\hat\ell\hat\kappa\cr
\Biggl\lbrace \sqrt{(2\ell_2+3)(\ell_2+1)}&
\Bigl\lbrace\matrix{\ell&1&\kappa\cr \ell_2&\ell_1&\ell_2+1}\Bigr\rbrace
\Bigl(\matrix{\ell_1&\ell&\ell_2+1\cr 0&0&0}\Bigr)\langle R_1(r)\vert
j_{\ell}(qr) ({d\over{dr}}-{\ell_2\over{r}})\vert R_2(r)\rangle\cr
-\sqrt{(2\ell_2-1)\ell_2}&
\Bigl\lbrace\matrix{\ell&1&\kappa\cr \ell_2&\ell_1&\ell_2-1}\Bigr\rbrace
\Bigl(\matrix{\ell_1&\ell&\ell_2-1\cr 0&0&0}\Bigr)\langle R_1(r)\vert
j_{\ell}(qr) ({d\over{dr}}+{{\ell_2+1}\over{r}})\vert R_2(r)\rangle
\Biggr\rbrace\cr}\eqno{4.6}$$
which we abbreviate as 
$$\sum_{i=1,2}del(i,\ell,\kappa,1,2)\langle R_1(r)\vert
j_{\ell}(qr) ~op(i)\vert R_2(r)\rangle\eqno{4.7}$$
where $op(1)={d\over{dr}}$ and $op(2)={1\over{r}}$. 

And the matrix element involving the operator $\nabla^2$: 
$$\eqalign{\sqrt{4\pi}
 \langle \ell_1\Vert \Bigl[ [Y^{(\ell)}
  \otimes\bigl[\vec\nabla\otimes & \vec\nabla\bigr]^{(J)}\Bigr]^{(\lambda )} 
  \Vert \ell_2\rangle=(-)^{(\ell_2+\lambda+J)}\hat \lambda\hat J
\hat \ell_1\hat\ell\cr
  \Bigl[ & 
  \Bigl\lbrace\matrix{\ell&J&\lambda\cr \ell_2&\ell_1&\ell_2+2}\Bigr\rbrace
  \Bigl\lbrace\matrix{1&1&J\cr \ell_2&\ell_2+2&\ell_2+1}\Bigr\rbrace
  \Bigl(\matrix{\ell_1&\ell&\ell_2+2\cr 0&0&0}\Bigr) \cr
  &~~~~~~~~~~\times \widehat{(\ell_2+2)}\sqrt{\ell_2+1}\sqrt{\ell_2+2}
  <R_1(r)\vert
  j_{\ell}(qr)\bigl({d\over{dr}}-{{\ell_2+1}\over{r}}\bigr)
  \bigl({d\over{dr}}-{\ell_2\over{r}}\bigr)\vert R_2(r)>\cr
  -&\Bigl\lbrace\matrix{\ell&J&\lambda\cr \ell_2&\ell_1&\ell_2}\Bigr\rbrace
  \Bigl\lbrace\matrix{1&1&J\cr \ell_2&\ell_2&\ell_2+1}\Bigr\rbrace
  \Bigl(\matrix{\ell_1&\ell&\ell_2\cr 0&0&0}\Bigr) \cr
  &~~~~~~~~~~\times \hat \ell_2(\ell_2+1)
  <R_1(r)\vert
  j_{\ell}(qr)\bigl({d\over{dr}}+{{\ell_2+2}\over{r}}\bigr)
  \bigl({d\over{dr}}-{\ell_2\over{r}}\bigr)\vert R_2(r)>\cr
  -&\Bigl\lbrace\matrix{\ell&J&\lambda\cr \ell_2&\ell_1&\ell_2}\Bigr\rbrace
  \Bigl\lbrace\matrix{1&1&J\cr \ell_2&\ell_2&\ell_2-1}\Bigr\rbrace
  \Bigl(\matrix{\ell_1&\ell&\ell_2\cr 0&0&0}\Bigr) \cr
  &~~~~~~~~~~\times\hat\ell_2\ell_2
  <R_1(r)\vert
  j_{\ell}(qr)\bigl({d\over{dr}}-{{\ell_2-1}\over{r}}\bigr)
  \bigl({d\over{dr}}+{{\ell_2+1}\over{r}}\bigr)\vert R_2(r)>\cr
  +&\Bigl\lbrace\matrix{\ell&J&\lambda\cr \ell_2&\ell_1&\ell_2-2}\Bigr\rbrace
  \Bigl\lbrace\matrix{1&1&J\cr \ell_2&\ell_2-2&\ell_2-1}\Bigr\rbrace
  \Bigl(\matrix{\ell_1&\ell&\ell_2-2\cr 0&0&0}\Bigr) \cr
  &~~~~~~~~~~\times \widehat{(\ell_2-2)}\sqrt{\ell_2-1}\sqrt{\ell_2}
  <R_1(r)\vert
  j_{\ell}(qr)\bigl({d\over{dr}}+{{\ell_2}\over{r}}\bigr)
  \bigl({d\over{dr}}+{{\ell_2+1}\over{r}}\bigr)\vert R_2(r)>
}\eqno{4.8}$$
Here we give the reduced matrix elements separately for $J=0$ and for
$J=2$ since they combine with different form factors. 
For $J=0$ the selection rules require
$\ell=\lambda$. Thus for $J=0$ we can abbreviate (4.8) as
$$\sum_{i=1,3}~ddel_0(i,\lambda,1,2)<R_1(r)\vert
  j_{\ell}(qr)op'(i)\vert R_2(r)>\eqno{4.9}$$
whereas for $J=2$ we abbreviate this as
$$\sum_{i=1,3}~ddel_2(i,\ell,\lambda,1,2)<R_1(r)\vert
  j_{\ell}(qr)op'(i)\vert R_2(r)>\eqno{4.10}$$
with $op'(1)={d^2\over{dr^2}}$, $op'(2)={1\over{r}}{d\over{dr}}$,
and $op'(3)={1\over{r^2}}$.

Furthermore, we note that
$${3\over{2}}\langle 1010|00\rangle
  \Bigl\lbrace\matrix{1&1&1\cr 1&1&0}\Bigr\rbrace
  ={1\over{\sqrt{12}}}\eqno{4.11}
$$
and
$${3\over{2}}\langle 1010|20\rangle
  \Bigl\lbrace\matrix{1&1&1\cr 1&1&2}\Bigr\rbrace
  {1\over{\sqrt{5}}}
 =\sqrt{{1\over{120}}}
 \eqno{4.12}$$

The matrix element of the $J=0$ piece can be written as
$$\eqalign{
<(1\bar 2)_{\lambda}\vert V^{R,L2}_{(J=0)}\vert (3\bar 4)_{\lambda}>=&
(-)^{k_1+k_4}{2\over{\pi}}
\int q^2dq\tilde V^{L20}(q)\cr
&{1\over{\sqrt{12}}}
\Bigl({\sqrt{4\pi}\over{\hat\lambda}}
<1\Vert\bigl(Y^{\lambda}\otimes\bigl[
\nabla\otimes\nabla\bigr]^{(0)}
\bigr)^{(\lambda)}\Vert 2>
{\sqrt{4\pi}\over{\hat\lambda}}<4\Vert Y^{\lambda}\Vert 3>\cr
&+{\sqrt{4\pi}\over{\hat\lambda}}<1\Vert Y^{\lambda}\Vert 2>
{\sqrt{4\pi}\over{\hat\lambda}}
<4\Vert\bigl(Y^{\lambda}\otimes\bigl[\nabla\otimes\nabla\bigr]^{(0)}
\bigr)^{(\lambda)}\Vert 3>\Bigr)}\eqno{4.13}$$
This matrix element contributes only in natural parity cases.
$$\eqalign{
<(1\bar 2)_{\lambda}\vert V^{R,L2}_{Z=2,(J=0)}\vert (3\bar 4)_{\lambda}>=&
{4\pi\over{2\lambda+1}}(-)^{k_1+k_4}{2\over{\pi}}
\int q^2dq\tilde V^{L20}(q)\cr
&\sum_{\ell=\lambda-1,.,\lambda+1}
(-)^{\ell+\lambda+1}
{1\over{3}}
<1\Vert (Y^{\ell}\nabla)^{\lambda}\Vert 2>
<4\Vert (Y^{\ell}\nabla)^{\lambda}\Vert 3>
}\eqno{4.14}$$

For the term with $J=1$, linear in r, the contributions are from
$Z_4$ and $Z_5$:
$$\eqalign{
<(1\bar 2)_{\lambda}\vert V^{R,L2}_{(J=1)}\vert (3\bar4)_{\lambda}>=&\cr
{1\over{\sqrt{3}}}{4\pi\over{2\lambda+1}}(-)^{(k_1+k_4)}
\sum_{\ell=\lambda-1,\lambda+1}&(-)^{(\ell+\lambda+1)/2}
\hat \ell <\ell0\lambda 0|10>{2\over{\pi}}
\int q^2dq\bar V^{L2,1}(q)\cr
&\Bigl(<1\Vert (Y^{\ell}\nabla)^{(\lambda)}\Vert 2>
       <4\Vert Y^{\lambda}\Vert 3> 
     + <1\Vert Y^{\lambda}\Vert 2>
       <4\Vert (Y^{\ell}\nabla)^{(\lambda)}\Vert 3>\Bigr)}\eqno{4.15}$$
This term contributes only in natural parity cases. We have to consider
$\ell=\lambda -1$ and $\ell=\lambda +1$.

Finally, the contributions from $J=2$ can be written 
$$\eqalign{
<(1\bar 2)_{\lambda}\vert V^{R,L2}_{(J=2)Z=1}\vert (3\bar
4)_{\lambda}>=
-{1\over{\sqrt{120}}}{4\pi\over{2\lambda+1}}&(-)^{(k_1+k_4)}
{2\over{\pi}}\int q^2dq \tilde V^{L2,2}(q)\cr
\sum_{\ell=\lambda-2,\lambda,\lambda+2}(-)^{(\ell+\lambda)/2}\hat \ell
<\ell 0\lambda 0\vert 20>\Bigl(&
<1\Vert \bigl[Y^{\ell}(\nabla\otimes\nabla)^{(2)}\bigr]^{(\lambda)}
\Vert 2> <4\Vert Y^{\lambda}\Vert 3>\cr
&+ <1\Vert Y^{\lambda}\Vert 2>
<4\Vert \bigl[Y^{\ell}(\nabla\otimes\nabla)^{(2)}\bigr]^{(\lambda)}
\Vert 3>\Bigr) }\eqno{4.16}$$
Again, this term contributes only in natural parity cases with
$\ell=\lambda -2$, $\ell=\lambda$, and $\ell=\lambda +2$.

The remaining contribution is
$$\eqalign{
<(1\bar 2)_{\lambda}\vert& V^{R,L2}_{(J=2)Z=2}\vert (3\bar
4)_{\lambda}>=\cr
-&{4\pi\over{2\lambda+1}}(-)^{(k_1+k_4)}
{2\over{\pi}}\int q^2dq \tilde V^{L2,2}(q)\cr
&\sum_{\ell_a=\lambda\pm 1}\sum_{\ell_b=\lambda\pm 1}
d_{lsq}(\ell_a,\ell_b,\lambda)
<1\Vert (Y^{\ell_a}\nabla)^{(\lambda)}\Vert 2>
<4\Vert (Y^{\ell_b}\nabla)^{(\lambda)}\Vert 3>\Bigr)
}\eqno{4.17}$$
with
$$d_{lsq}(\ell_a,\ell_b,\lambda)=\sqrt{{1\over{6}}}(-)^{(\ell_a+\ell_b)/2}
\hat \ell_a\hat \ell_b<\ell_a0\ell_b0\vert20>
\Bigl\lbrace\matrix{1&1&2\cr\ell_a&\ell_b&\lambda}\Bigr\rbrace\eqno{4.18}$$
This term contributes in both, natural and unnatural parity cases.
We have to consider the three cases with $\ell_a=\ell_b=\lambda -1$,
$\lambda$, and $\lambda +1$ as well as the two cases:
$\ell_a=\lambda -1,\ell_b=\lambda+1$ and
$\ell_a=\lambda +1,\ell_b=\lambda-1$.

With our abbreviations the reduced matrix elements can be
written as
$$ (-)^{k_1}{\sqrt{4\pi}\over{\hat\lambda}} \langle 1\Vert (Y^{\ell}
\vec\nabla )^{\lambda}\Vert 2\rangle=
\sum_{i=1,2}f(\lambda,1,2)del(i,\ell,\lambda,1,2)
\langle R_1(r)\vert
j_{\ell}(qr)~ op(i)\vert R_2(r)\rangle\eqno{4.19}$$
$${\sqrt{4\pi}\over{\hat \lambda}}(-)^{k_1}
<1\Vert  \bigl[Y^{\lambda}(\nabla\otimes\nabla)^{(0)}\bigr]^{(\lambda)}
\Vert 2>=\sum_{i=1,3}f(\lambda,1,2)ddel_0(i,\lambda,1,2)
\langle R_1(r)\vert
j_{\ell}(qr)~ op'(i)\vert R_2(r)\rangle\eqno{4.20}$$
and
$${\sqrt{4\pi}\over{\hat \lambda}}(-)^{k_1}
<1\Vert  \bigl[Y^{\ell}(\nabla\otimes\nabla)^{(2)}\bigr]^{(\lambda)}
\Vert 2>=\sum_{i=1,3}f(\lambda,1,2)ddel_2(i,\ell,\lambda,1,2)
\langle R_1(r)\vert
j_{\ell}(qr)~ op'(i)\vert R_2(r)\rangle\eqno{4.21}$$

\beginsection{5. Terms $\sigma\sigma$ and (LL)$\sigma\sigma$}

Such terms can be obtained from the previous terms such as the central
interaction or the $L^2$-interaction by adding the
$\sigma_1\sigma_2$ term. This results in
$$
V=V^{L2\sigma}(r_{12})\bigl[ L\odot L\bigr]\bigl[\vec\sigma_1\odot
  \vec\sigma_2\bigr]
  \eqno{5.1}$$
We can obtain the separated interaction immediately from the separated
form for the $L^2$ term via
$$\bigl[F^{(J)}(1)\odot
G^{(J)}(2)\bigr]\bigl[\vec\sigma_1\odot\vec\sigma_2
\bigr]=\sum_k(-)^{(k+J+1)}\Bigl[\bigl[F^{(J)}(1)\otimes\vec\sigma_1\bigr]
^{(k)}\odot\bigl[G^{(J)}(2)\otimes\vec\sigma_2\bigr]^{(k)}\Bigr]
\eqno{5.2}$$

Thus, we can immediately take the results of the previous sections and
write the matrix element for the $\sigma$-interaction as
$$\eqalign{
<(1\bar 2)_{\lambda}\vert V^{R,\sigma}_{(J=0)}\vert (3\bar 4)_{\lambda}>=&
{4\pi\over{2\lambda+1}}(-)^{k_1+k_4}{2\over{\pi}}
\int q^2dq\tilde V^{\sigma}(q)\cr
&\sum_{\ell=\lambda-1,.,\lambda+1}(-)^{(\ell+\lambda+1)}
<1\Vert (Y^{\ell}\sigma)^{\lambda}\Vert 2>
<4\Vert (Y^{\ell}\sigma)^{\lambda}\Vert 3>}\eqno{5.3}$$

Similarly we write the contributions to the matrix elements of the 
$L^2\sigma$-interaction for $J=0$ as
$$\eqalign{
<(1\bar 2)_{\lambda}\vert V^{R,L2\sigma}_{(J=0)}\vert (3\bar 4)_{\lambda}>=
{1\over{\sqrt{12}}}&{4\pi\over{2\lambda+1}}(-)^{k_1+k_4}{2\over{\pi}}
\int q^2dq\tilde V^{L20\sigma}(q)
\sum_{\ell=\lambda-1,.,\lambda+1}(-)^{(\ell+\lambda+1)}\cr
\Bigl(&<1\Vert\Bigl[\bigl(Y^{\ell}\otimes\bigl[
\nabla\otimes\nabla\bigr]^{(0)}
\bigr)^{(\ell)}\sigma\Bigr]^{\lambda}\Vert 2><4\Vert
(Y^{\ell}\sigma)^{\lambda}\Vert 3>\cr
&+<1\Vert (Y^{\ell}\sigma)^{\lambda}\Vert 2>
<4\Vert\Bigl[\bigl(Y^{\ell}\otimes\bigl[\nabla\otimes\nabla\bigr]^{(0)}
\bigr)^{\ell}\sigma\Bigr]^{(\lambda)}\Vert 3>\Bigr)}\eqno{5.4}$$
We include three cases: $\ell=\lambda -1$, $\ell=\lambda$, and
$\ell-\lambda +1$.
$$\eqalign{
<(1\bar 2)_{\lambda}\vert V^{R,L2\sigma}_{Z=2(J=0)}\vert (3\bar 4)_{\lambda}>=&
{1\over{3}}{4\pi\over{2\lambda+1}}(-)^{k_1+k_4}{2\over{\pi}}
\int q^2dq\tilde V^{L20\sigma}(q)\cr
&\sum_{\kappa}
\sum_{\ell=\kappa-1,\kappa+1}
(-)^{(\ell+\lambda)}
<1\Vert \Bigl[(Y^{\ell}\nabla)^{\kappa}\sigma\Bigr]^{(\lambda)}\Vert 2>
<4\Vert \Bigl[(Y^{\ell}\nabla)^{\kappa}\sigma\Bigr]^{(\lambda)}\Vert 3>
\Bigr\rbrace}\eqno{5.5}$$
In this term $\ell$ can take the values of $\lambda-2$, $\lambda-1$,
$\lambda$, $\lambda+1$, and $\lambda+2$.

For the term with $J=1$, linear in r the contributions are from
$Z_4$ and $Z_5$:
$$\eqalign{
&<(1\bar 2)_{\lambda}\vert V^{R,L2\sigma}_{(J=1)}\vert (3\bar4)_{\lambda}>=\cr
&\sum_{\kappa}
\sum_{\ell=\kappa-1,\kappa+1}(-)^{(\ell+\kappa+1)/2}
(-)^{(\kappa+\lambda+1)}
{1\over{\sqrt{3}}}{4\pi\over{2\lambda+1}}(-)^{(k_1+k_4)}
\hat \ell <\ell0\kappa 0|10>{2\over{\pi}}
\int q^2dq\bar V^{L2\sigma,1}(q)\cr
&\Bigl(<1\Vert \bigl[(Y^{\ell}\nabla)^{(\kappa)}\sigma\bigr]^{(\lambda)}\Vert 2>
       <4\Vert (Y^{\kappa}\sigma)^{(\lambda)}\Vert 3>
     + <1\Vert (Y^{\kappa}\sigma)^{(\lambda)}\Vert 2>
       <4\Vert \bigl[(Y^{\ell}\nabla)^{(\kappa)}\sigma\bigr]^{(\lambda)}\Vert 3>
 \Bigr)}\eqno{5.6}$$
Here $\kappa$ can be $\lambda -1$, $\lambda$, and $\lambda +1$.
For each case $\ell$ can be $\kappa +1$ and $\kappa -1$.

Finally, the contributions from $J=2$ can be written 
$$\eqalign{
<(1\bar 2)_{\lambda}\vert V^{R,L2\sigma}_{(J=2)Z=1}\vert (3\bar
4)_{\lambda}>=\sum_{\kappa}(-)&^{(\kappa+\lambda)}
{1\over{\sqrt{120}}}{4\pi\over{2\lambda+1}}(-)^{(k_1+k_4)}
{2\over{\pi}}\int q^2dq \tilde V^{L2\sigma,2}(q)\cr
\sum_{\ell=\kappa-2,\kappa,\kappa+2}(-)^{(\ell+\kappa)/2)}\hat \ell
<\ell 0\kappa 0\vert 20>\Bigl( &
<1\Vert
\Bigl[\bigl[Y^{\ell}(\nabla\otimes\nabla)^{(2)}\bigr]^{(\kappa)}
\sigma\Bigr]^{(\lambda)}
\Vert 2> <4\Vert (Y^{\kappa}\sigma)^{(\lambda)}\Vert 3>\cr
+& <1\Vert (Y^{\kappa}\sigma)^{(\lambda)}\Vert 2>
<4\Vert
\Bigl[\bigl[Y^{\ell}(\nabla\otimes\nabla)^{(2)}\bigr]^{(\kappa)}
\sigma\Bigr]^{(\lambda)}\Vert 3>\Bigr)}\eqno{5.7}$$
Again, $\kappa$ can take the values $\lambda -1$, $\lambda$,
and $\lambda+1$. For each case $\ell$ can be $\kappa$, $\kappa-2$,
and $\kappa+2$.
$$\eqalign{
<(1\bar 2)_{\lambda}\vert& V^{R,L2\sigma}_{(J=2)Z=2}\vert (3\bar
4)_{\lambda}>=\sum_{\kappa}(-)^{(\kappa+\lambda)}
{1\over{\sqrt{120}}}{4\pi\over{2\lambda+1}}(-)^{(k_1+k_4)}
{2\over{\pi}}\int q^2dq \tilde V^{L2\sigma,2}(q)\cr
+&\sum_{\ell_a=\kappa\pm 1}\sum_{\ell_b=\kappa\pm 1}
dl(\ell_a,\ell_b,\kappa)
<1\Vert \bigl[(Y^{\ell_a}\nabla)^{(\kappa)}\sigma\bigr]^
{(\lambda)}\Vert 2>
<4\Vert \bigl[(Y^{\ell_b}\nabla)^{(\kappa)}\sigma\bigr]^
{(\lambda)}\Vert 3>\Bigr)
}\eqno{5.8}$$

For the reduced one-body matrix elements containing $\sigma$ we use
([3] eq. 7.1.5) 
$${\sqrt{4\pi}\over{\hat \lambda}}(-)^{k_1}
<(\ell_1s_1)j_1\Vert (G^{(\ell)}\sigma)^{(\lambda)}\Vert
(\ell_2s_2)j_2>=(-)^{k_1}
\hat j_1\hat j_2 \sqrt{6}
\Biggl\lbrace\matrix{\ell_1&\ell_2&\ell\cr 1/2&1/2&1\cr j_1&j_2&\lambda}
\Biggr\rbrace \sqrt{4\pi}
<\ell_1\Vert G^{(\ell)}\Vert \ell_2>\eqno{5.9}$$
we abbreviate this as
$$\sum_\ell g(\ell,k,1,2)\sqrt{4\pi}<\ell_1\Vert G^{(\ell)}\Vert
\ell_2>\eqno{5.10}$$
With our abbreviations the reduced matrix elements can be
written as
$$(-)^{k_1}{\sqrt{4\pi}\over{\hat\lambda}} \langle 1\Vert (Y^{\ell}\vec\sigma)^
  {\lambda} \Vert 2\rangle=
  g(\ell,\lambda,1,2)y(\ell,1,2)
  ~\langle R_1(r)\vert j_{\ell}(qr)\vert R_2(r)\rangle\eqno{5.11}$$
$${\sqrt{4\pi}\over{\hat\lambda}}(-)^{k_1} \langle 1\Vert \bigl[ [Y^{\ell}
  \otimes\vec\nabla]^{\kappa}\otimes\vec\sigma\bigr]^{\lambda} 
  \Vert 2\rangle=
  \sum_{i=1,2}g(\kappa,\lambda,1,2)
  del(i,\ell,\kappa,1,2)\langle R_1(r)\vert
  j_{\ell}(qr)~ op(i)\vert R_2(r)\rangle\eqno{5.12}$$
$${\sqrt{4\pi}\over{\hat \lambda}}(-)^{k_1}
  \langle 1\Vert \Bigl[ \bigl[Y^{\kappa}(\nabla\otimes\nabla)^{(0)}\bigr]^{(\kappa)}
  \sigma\Bigr]^{(\lambda)}
  \Vert 2\rangle=
  \sum_{i=1,3}g(\kappa,\lambda,1,2)ddel_0(i,\kappa,1,2)
  \langle R_1(r)\vert
  j_{\ell}(qr)~ op'(i)\vert R_2(r)\rangle\eqno{5.13}$$
and
$${\sqrt{4\pi}\over{\hat \lambda}}(-)^{k_1}
\langle 1\Vert  \Bigl[\bigl[Y^{\ell}(\nabla\otimes\nabla)^
{(2)}\bigr]^{(\kappa)}\sigma\Bigr]^{(\lambda)}
\Vert 2\rangle=\sum_{i=1,3}g(\kappa,\lambda,1,2)ddel_2(i,\ell,\kappa,1,2)
\langle R_1(r)\vert
j_{\ell}(qr)~ op'(i)\vert R_2(r)\rangle\eqno{5.14}$$
\vfill

\beginsection{6. Spin-orbit (LS)}

$$\eqalign{V=&V^{LS}(r_{12})\bigl[L^{(1)}\odot S^{(1)}\bigr]=
   -\sqrt{3}V^{LS}(r_{12})\bigl[L^{(1)}\otimes S^{(1)}\bigr]^{(0)}\cr
  =&-r_{12}V^{LS}(r_{12})\sqrt{2\pi}\Bigl[\bigl[Y^{(1)}(\hat
r_{12})\otimes(\vec\nabla_1-\vec\nabla_2)^{(1)}\bigr]^{(1)}
\otimes S^{(1)}\Bigr]^{(0)}}\eqno{6.1}
$$
We introduce the form-factor of the LS-interaction:
$$\tilde V^{LS}(q)=\int r_{12}^3V^{LS}(r_{12})j_1(qr_{12})dr_{12}
\eqno{6.2}$$
and write using eq.(2.2) with $J=1$
$$\eqalign{ V=&-
  \sqrt{2\pi}  \sum_{\ell_1,\ell_2}{2\over{\pi}}\int
  q^2dq\tilde V^{LS}(q)j_{\ell_1}(qr_1)j_{\ell_2}(qr_2)\cr
  &~~~\times \hat \ell_1 \hat \ell_2 {1\over{\sqrt{3\cdot 4\pi}}}
  <\ell_1 0\ell_2 0|10> (i)^{\ell_1}(-i)^{\ell_2+1}\cr
  &~~~\times 4\pi\biggl[\Bigl[\bigl[ Y^{(\ell_1)}(\hat r_1)\otimes Y^{(\ell_2)}
  (\hat r_2)\bigr]^{(1)}\otimes (\vec\nabla_1-\vec\nabla_2)^{(1)}\Bigr]
  ^{(1)}\otimes S^{(1)}\biggr]^{(0)}}\eqno{6.3}
  $$
Or, combining factors
$$\eqalign{ V=&\sum_{\ell_1,\ell_2}(-)^{(\ell_1-\ell_2+1)/2}
   \sqrt{{1\over{24}}} {2\over{\pi}}\int
  q^2dq\tilde V^{LS}(q)j_{\ell_1}(qr_1)j_{\ell_2}(qr_2)\cr
  &~~~\times \hat \ell_1 \hat \ell_2 
  <\ell_1 0\ell_2 0|10> \cr
  &~~~\times 4\pi\biggl[\Bigl[\bigl[ Y^{(\ell_1)}(\hat r_1)\otimes Y^{(\ell_2)}
  (\hat r_2)\bigr]^{(1)}\otimes (\vec\nabla_1-\vec\nabla_2)^{(1)}\Bigr]
  ^{(1)}\otimes (\sigma_1+\sigma_2)^{(1)}\biggr]^{(0)}}\eqno{6.4}
  $$
Inserting $S$ from above, results in four terms as represented in the
last line of the previous equation. They must be recoupled and brought
into the form $F^{(k)}(1)\odot G^{(k)}(2)$. These terms are:
$$\eqalign{
Z_1=&\biggl[\Bigl[\bigl[ Y^{\ell_1}(1)\otimes Y^{\ell_2}(2)\bigr]^{(1)}
\otimes \vec\nabla_1^{(1)}\Bigr]^{(1)}\otimes \sigma_1^{(1)}\biggr]
^{(0)}\cr
~&=-\sum_k\sqrt{3}\hat k
\Bigl\lbrace\matrix{\ell_2&\ell_1&1\cr 1&1&k}\Bigr\rbrace
\biggl[Y^{\ell_2}(2)\otimes \Bigl[\bigl[ Y^{\ell_1}(1)\otimes
\vec\nabla_1\bigr]^{(k)} \otimes \sigma_1\Bigr]^{(\ell_2)}\biggr]^{(0)}\cr
~&=(-)^{(\ell_2+1)}\sum_k\sqrt{3}{{\hat k}\over{\hat \ell_2}}
\Bigl\lbrace\matrix{\ell_2&\ell_1&1\cr 1&1&k}\Bigr\rbrace
\biggl[Y^{\ell_2}(2)\odot \Bigl[\bigl[ Y^{\ell_1}(1)\otimes
\vec\nabla_1\bigr]^{(k)} \otimes \sigma_1\Bigr]^{(\ell_2)}\biggr]\cr
Z_2=&(-)\biggl[\Bigl[\bigl[  Y^{\ell_1}(1)\otimes Y^{\ell_2}(2)\bigr]^{(1)}
\otimes \vec\nabla_2^{(1)}\Bigr]^{(1)}\otimes \sigma_1^{(1)}\biggr]
^{(0)}\cr
~&=(-)^{\ell_2}\sqrt{3}\sum_k
\Bigl\lbrace\matrix{\ell_2&\ell_1&1\cr 1&1&k}\Bigr\rbrace
\Bigl[\bigl[Y^{\ell_1}(1)\otimes\sigma_1\bigr]^{(k)}\odot
      \bigl[Y^{\ell_2}(2)\otimes\nabla_2\bigr]^{(k)}\Bigr]\cr
Z_3=&\biggl[\Bigl[\bigl[ Y^{\ell_1}(1)\otimes Y^{\ell_2}(2)\bigr]^{(1)}
\otimes \vec\nabla_1^{(1)}\Bigr]^{(1)}\otimes \sigma_2^{(1)}\biggr]
^{(0)}\cr
~&=(-)^{\ell_2}\sqrt{3}\sum_k
\Bigl\lbrace\matrix{\ell_2&\ell_1&1\cr 1&1&k}\Bigr\rbrace
\Bigl[\bigl[ Y^{\ell_2}(2)\otimes\sigma_2\bigr]^{(k)}\odot
\bigl[Y^{\ell_1}(1)\otimes\nabla_1)\bigr]^{(k)}\Bigr]\cr
Z_4=&(-)\biggl[\Bigl[\bigl[  Y^{\ell_1}(1)\otimes Y^{\ell_2}(2)\bigr]^{(1)}
\otimes \vec\nabla_2^{(1)}\Bigr]^{(1)}\otimes \sigma_2^{(1)}\biggr]
^{(0)}\cr
~&=(-)^{(\ell_2+1)}\sum_k\sqrt{3}{{\hat k}\over{\hat \ell_1}}
\Bigl\lbrace\matrix{\ell_2&\ell_1&1\cr 1&1&k}\Bigr\rbrace
\biggl[Y^{\ell_1}(1)\odot \Bigl[\bigl[ Y^{\ell_2}(2)\otimes
\vec\nabla_2\bigr]^{(k)} \otimes \sigma_2\Bigr]^{(\ell_1)}\biggr]
}\eqno{6.5}$$
 by introducing
$$d_{so}(\ell_1,\ell_2,\lambda)={-1\over{\sqrt{8}}}\hat \ell_1\hat\lambda
<\ell_1~0~\ell_2~0|1~0>(-)^{(\ell_1+\ell_2+1)/2}
\Bigl\lbrace\matrix{1&1&1\cr \ell_1&\ell_2&\lambda}\Bigr\rbrace
\eqno{6.6}$$
we write the matrix element as
$$\eqalign{
<(1,\bar 2)_{\lambda}|V^{R,LS}|(3,\bar 4)_{\lambda}>=&
{4\pi\over{2\lambda +1}}(-)^{(k_1+k_4)}\sum_
{\ell=\lambda\pm 1} {2\over{\pi}}\int q^2dq~\tilde V^{LS}(q)\cr
\biggl(\sum_k~&d_{so}(\ell,\lambda,k)
<1\Vert\Bigl[\bigl[ Y^{\ell}\otimes
\vec\nabla\bigr]^{(k)} \otimes \sigma\Bigr]^{(\lambda)}\Vert 2>
<4\Vert Y^{\lambda}\Vert 3>\cr
+\sum_k~&d_{so}(\ell,\lambda,k)
<4\Vert\Bigl[\bigl[ Y^{\ell}\otimes
\vec\nabla\bigr]^{(k)} \otimes \sigma\Bigr]^{(\lambda)}\Vert 3>
<1\Vert Y^{\lambda}\Vert 2>\cr
-&d_{so}(\ell,\lambda,\lambda)
<1\Vert (Y^{\ell}\sigma)^{(\lambda)}\Vert 2>
 <4\Vert (Y^{\lambda}\nabla )^{(\lambda)}\Vert 3>\cr
-&d_{so}(\ell,\lambda,\lambda)
<4\Vert (Y^{\ell}\sigma)^{(\lambda)}\Vert 3>
 <1\Vert (Y^{\lambda}\nabla )^{(\lambda)}\Vert 2>\cr
-&d_{so}(\ell,\lambda,\lambda)
<1\Vert (Y^{\lambda}\sigma)^{(\lambda)}\Vert 2>
 <4\Vert (Y^{\ell}\nabla )^{(\lambda)}\Vert 3>\cr
-&d_{so}(\ell,\lambda,\lambda)
<4\Vert (Y^{\lambda}\sigma)^{(\lambda)}\Vert 3>
 <1\Vert (Y^{\ell}\nabla )^{(\lambda)}\Vert 2>\biggr)
}\eqno{6.7}$$
All the reduced one body matrix elements have been defined
before.
\vskip 0.2in

\beginsection{7. Tensor interaction}

We write the tensor operator $S_{12}$ as
$$S_{12}=3(\vec\sigma_1\hat r_{12})(\vec\sigma_2\hat r_{12})-\vec
\sigma_1\vec\sigma_2\eqno{7.1}$$
Recoupling this, $S_{12}$ can be written as coupled tensor operator
$$S_{12}=3\sum_J\hat J\Bigl[\bigl[\sigma_1\otimes\sigma_2\bigr]
^{(J)}\otimes\bigl[\hat r_{12}\otimes\hat r_{12}\bigr]^{(J)}\Bigr]
^{(0)}+\sqrt{3}\bigl[\sigma_1\otimes\sigma_2\bigr]^{(0)}\eqno{7.2}$$
with only $J=0$ and $J=2$ contributing. In this sum the last term
is cancelled against the $J=0$ term in the sum. The remaining
term is
$$S_{12}=\sqrt{4\pi}~3<1010\vert 20>\Bigl[Y^2(\hat r_{12})\otimes\bigl[
\sigma_1\otimes\sigma_2\bigr]^{(2)}\Bigr]^{(0)}\eqno{7.3}$$
Separating $\hat r_{12}$ and evaluating the Clebsch-Gordan coefficient
($3<1010|20>=\sqrt{6}$)
finally gives
$$\eqalign{
<(1,\bar 2)_{\lambda}|V^{R,T}|(3,\bar 4)_{\lambda}>=&
{4\pi\over{2\lambda +1}}(-)^{(k_1+k_4)}\sum_{\ell_a}
\sum_{\ell_b}
{2\over{\pi}}\int q^2dq~\tilde V^{T}(q)\cr
\sqrt{6}\hat \ell_a\hat\ell_b<\ell_a0\ell_b0|20>(-)^{(\ell_a
+\ell_b)/2}&
\Bigl\lbrace\matrix{\ell_a&\ell_b&2\cr 1&1&\lambda}\Bigr\rbrace
<1\Vert (Y^{\ell_a}\sigma)^{(\lambda)}\Vert 2>
<4\Vert (Y^{\ell_b}\sigma)^{(\lambda)}\Vert 3>}\eqno{7.4}$$
where the form factor is
$$\tilde V^{T}(q)=\int V^T(r_{12})j_2(qr_{12})r_{12}^2dr_{12}\eqno{7.5}$$
\vfill

\beginsection{8. Term (LS)$^2$}

This term can be brought into a similar form as the some 
of the interactions already discussed. We write
$$V=V^{LS2}(r_{12})\bigl[\vec L \odot \vec S \bigr] 
\bigl[ \vec L \odot \vec S \bigr]=
V^{LS2}(r_{12})\sum_j\hat j\Bigl[\bigl[L^{(1)}\otimes 
L^{(1)}\bigr]^{(j)}\otimes \bigl[S^{(1)}\otimes
  S^{(1)}\bigr]^{(j)}\Bigr]^{(0)}\eqno{8.1}$$
Here $j$ can assume the values 0, 1 and 2.
While the terms with $j=0,1$ can immediately be deduced from the previous
cases, the term with $j=2$ needs to be evaluated new. 

For $j=0$ we write
$$\bigl[\vec L\otimes\vec L\bigr]^{(0)}\otimes\bigl[
\vec S\otimes \vec S\bigr]^{(0)}=
{1\over{3}}\bigl[\vec L\odot\vec L\bigr]\bigl[\vec S\odot
\vec S\bigr]=
{1\over{6}}\bigl[\vec L\odot\vec L\bigr] (3+\vec\sigma_1\cdot\vec\sigma_2)
\eqno{8.2}$$
Thus this part can be added immediately to the previously evaluated
interactions by using $V^{L2,eff}=V^{L2}+{1\over{2}}V^{LS2}$ and
$V^{L2\sigma,eff}=V^{L2\sigma}+{1\over{6}}V^{LS2}$.

For $j=1$ we write
$$\sqrt{3}\Bigl[\bigl[L^{(1)}\otimes 
L^{(1)}\bigr]^{(1)}\otimes \bigl[S^{(1)}\otimes
  S^{(1)}\bigr]^{(1)}\Bigr]^{(0)}=-{1\over{2}}
  \bigl[\vec L\odot\vec S\bigr]\eqno{8.3}$$
This contribution is incorporated by substituting
$V^{LS,eff}=V^{LS}-{1\over{2}}V^{LS2}$.

Here it remains to evaluate only the matrix elements of
$$V=V^{LS2}(r_{12})\sqrt{5}\Bigl[\bigl[L^{(1)}\otimes L^{(1)}\bigr]^{(2)}
\otimes \bigl[S^{(1)}\otimes
S^{(1)}\bigr]^{(2)}\Bigr]^{(0)}\eqno{8.4}$$
Using eq.(2.4) we write
$$ \bigl[S^{(1)}\otimes S^{(1)}\bigr]^{(2)}=
{1\over{4}}\Bigl( \bigl[\vec \sigma_1\otimes \vec\sigma_1\bigr]^{(2)}
+2\bigl[\vec\sigma_1\otimes\vec\sigma_2\bigr]^{(2)}
+\bigl[\vec\sigma_2\otimes\vec\sigma_2\bigr]^{(2)}\Bigr)$$
Noting that one-body matrix elements vanish for the operator
$ \bigl[ \sigma_1\otimes \sigma_1\bigr]^{(2)}$ we write the remaining
terms as
$$V=V^{LS2}(r_{12}){\sqrt{5}\over{2}}\Bigl[\bigl[L^{(1)}\otimes 
L^{(1)}\bigr]^{(2)}\otimes \bigl[\sigma_1^{(1)}\otimes
\sigma_2^{(1)}\bigr]^{(2)}\Bigr]^{(0)}\eqno{8.5}$$
We assume that similar to eq.(4.4) the operator $\bigl[ L\otimes L\bigr]
^{(2)}$ can be written as a tensor product of variables $\vec r_1$
and $\vec r_2$ as $\bigl[ U^{(\kappa_1)}(1)\otimes 
W^{(\kappa_2)}(2)\bigr]
^{(2)}$ Thus we can write
$$\eqalign{
\Bigl[\bigl[\vec L\otimes &\vec L\bigr]^{(2)}
\otimes\bigl[
\vec\sigma_1\otimes\vec\sigma_2\bigr]^{(2)}\Bigr]^{(0)}=
\Bigl[\bigl[ U^{(\kappa_1)}(1)\otimes W^{(\kappa_2)}(2)\bigr]
\otimes\bigl[
\vec\sigma_1\otimes\vec\sigma_2\bigr]^{(2)}\Bigr]^{(0)}
\cr &=\sqrt{5}\sum_k(-)^{(k_2+1)}
  \Bigl\lbrace\matrix{1&1&2\cr \kappa_1&\kappa_2&k}\Bigr\rbrace
  \Bigl[\bigl[U^{(\kappa_1)}(1)\otimes \vec\sigma_1\bigr]^{(k)}\odot
  \bigl[W^{(\kappa_2)}(2)\otimes \vec\sigma_2\bigr]^{(k)}\Bigr]^{(0)}
  }\eqno{8.6}$$
We now have to factorize the operator $\bigl[ L\otimes 
L\bigr]^{(2)}$ using eq.(1.10) with $j=2$. Again, as they
combine with different form factors, we list the three cases
separately namely: (a) $J_r=0,J_p=2$, (b) $J_r=2,J_p=0$, and
(c) $J_r=2,j_p=2$.
Numerically evaluating the 9j-symbol and CG-coefficient we find
$$\eqalign{
{\sqrt{5}\over{2}}\bigl[\vec L\otimes\vec L\bigr]^{(2)} 
=\sqrt{4\pi}\Bigl({\sqrt{5}\over{12}}
r^2\Bigl[&Y^{(0)}(\hat r)\otimes\bigl[\vec\nabla\otimes
\vec\nabla\bigr]^{(2)}\Bigr]^{(2)}
-{\sqrt{2}\over{24}}r^2\Bigl[Y^{(2)}(\hat r)
\otimes\bigl[\vec\nabla\otimes\vec\nabla\bigr]^{(0)}\Bigr]^{(2)}\cr
+{\sqrt{14}\over{12}}r^2\Bigl[&Y^{(2)}(\hat r)
\otimes\bigl[\vec\nabla\otimes
\vec\nabla\bigr]^{(2)}\Bigr]^{(2)}
-{1\over{4}}\sqrt{{5\over{3}}}r\bigl[Y^{(1)}(\hat r)\otimes
\vec\nabla\bigr]^{(2)}\Bigr)}\eqno{5.7}$$

In the following we list all the eleven terms that contribute
to $\bigl[L\otimes L\bigr]^{(2)}$ in their recoupled forms:
$$\eqalign{
Z_1=&\Bigl[\bigl[Y^{\ell_1}(1)\otimes Y^{\ell_2}(2)\bigr]^{(2)}
\otimes \bigl[\nabla_1\otimes \nabla_1\bigr]^{(0)}\Bigr]^{(2)}\cr
&=\Bigl[\bigl[ Y^{\ell_1}\otimes \bigl[\nabla_1\otimes\nabla_1\bigr]
^{(0)}\bigr]^{(\ell_1)}\otimes Y^{\ell_2}(2)\Bigr]^{(2)}\cr
Z_2=&-2\Bigl[\bigl[Y^{\ell_1}(1)\otimes Y^{\ell_2}(2)\bigr]^{(2)}
\otimes \bigl[\nabla_1\otimes\nabla_2\bigr]^{(0)}\Bigr]^{(2)}\cr
&={2\over{\sqrt{3}}}\sum_{L_1,L_2}(-)^{(L_2+\ell_1)}
\hat L_1\hat L_2
\Bigl\lbrace\matrix{L_1&L_2&2\cr\ell_2&\ell_1&1}\Bigr\rbrace
\Bigl[\bigl[Y^{\ell_1}(1)\nabla_1\bigr]^{(L_1)}\otimes
\bigl[Y^{\ell_2}(2)\nabla_2\bigr]^{(L_2)}\Bigr]^{(2)}\cr
Z_3=&\Bigl[Y^{\ell_1}(1)\otimes\Bigl[Y^{\ell_2}(2)\otimes
\bigl[\nabla_2\otimes\nabla_2\bigr]^{(0)}\Bigr]^{(\ell_2)}\Bigr]^{(2)}\cr
Z_4=&\Bigl[\bigl[Y^{\ell_1}(1)\otimes Y^{\ell_2}(2)\bigr]^{(0)}
\otimes \bigl[\nabla_1\otimes \nabla_1\bigr]^{(2)}\Bigr]^{(2)}\cr
&=\sum_k\delta_{\ell_1,\ell_2}
{\hat k\over{\hat \ell_1\sqrt{5}}}
\Bigl[\bigl[Y^{\ell_1}(1)\otimes\bigl[\nabla_1\otimes\nabla_1\bigr]
^{(2)}\Bigr]^{(k)}\otimes Y^{\ell_2}(2)\Bigr]^{(2)}\cr
Z_5=&-2\Bigl[\bigl[Y^{\ell_1}(1)\otimes Y^{\ell_2}(2)\bigr]^{(0)}
\otimes \bigl[\nabla_1\otimes\nabla_2\bigr]^{(2)}\Bigr]^{(2)}\cr
&=2\delta_{\ell_1,\ell_2}\sum_{L_1,L_2}(-)^{(L_1+\ell_1)}
{{\hat L_1\hat L_2}\over{\ell_1}}
\Bigl\lbrace\matrix{1&1&2\cr L_2&L_1&\ell_1}\Bigr\rbrace
\Bigl[\bigl[Y^{\ell_1}(1)\nabla_1\bigr]^{(L_1)}\otimes
\bigl[Y^{\ell_2}(2)\nabla_2\bigr]^{(L_2)}\Bigr]^{(2)}\cr
Z_6=&\sum_k(-)^{(k+\ell_2)}\delta_{\ell_1,\ell_2}
{\hat k\over{\hat \ell_1\sqrt{5}}}
\Bigl[ Y^{\ell_1}(1)\otimes
\bigl[Y^{\ell_2}(2)\otimes\bigl[\nabla_2\otimes\nabla_2\bigr]
^{(2)}\Bigr]^{(k)}\Bigr]^{(2)}\cr
Z_7=&\Bigl[\bigl[Y^{\ell_1}(1)\otimes Y^{\ell_2}(2)\bigr]^{(2)}
\otimes \bigl[\nabla_1\otimes \nabla_1\bigr]^{(2)}\Bigr]^{(2)}\cr
&=\sqrt{5}\sum_k(-)^{(\ell_1+k)}\hat k\Bigl\lbrace\matrix{
2&2&2\cr\ell_2&\ell_1&k}\Bigr\rbrace
\Bigl[\bigl[Y^{\ell_1}(1)\otimes\bigl[\nabla_1\otimes\nabla_1
\bigr]^{(2)}\bigr]^{(k)}\otimes Y^{\ell_2}(2)\Bigr]^{(2)}\cr
Z_8=&-2\Bigl[\bigl[Y^{\ell_1}(1)\otimes Y^{\ell_2}(2)\bigr]^{(2)}
\otimes \bigl[\nabla_1\otimes\nabla_2\bigr]^{(2)}\Bigr]^{(2)}\cr
&=-10\sum_{L_1,L_2}\hat L_1\hat L_2
\Biggl\lbrace\matrix{\ell_1&\ell_2&2\cr 1&1&2\cr
L_1&L_2&2}\Biggr\rbrace
\Bigl[\bigl[Y^{\ell_1}(1)\nabla_1\bigr]^{(L_1)}\otimes
\bigl[Y^{\ell_2}(2)\nabla_2\bigr]^{(L_2)}\Bigr]^{(2)}\cr
Z_9=&\sqrt{5}\sum_k\hat k\Bigl\lbrace\matrix{
2&2&2\cr\ell_2&\ell_1&k}\Bigr\rbrace
\Bigl[Y^{\ell_1}(1)\otimes\Bigl[Y^{\ell_2}(2)\otimes\bigl[\nabla_2
\otimes\nabla_2\bigr]^{(2)}\bigr]^{(k)}\Bigr]^{(2)}\cr
Z_{10}=&\Bigl[\bigl[Y^{\ell_1}\otimes Y^{\ell_2}\bigr]^{(1)}
\otimes \nabla_1\Bigr]^{(2)}\cr
&=\sum_k (-)^{(\ell_2+k)}\sqrt{3}\hat k
\Bigl\lbrace\matrix{1&2&1\cr\ell_2&\ell_1&k}\Bigr\rbrace
\Bigl[\bigl[Y^{\ell_1}(1)\nabla_1\bigr]^{(k)}\otimes Y^{\ell_2}(2)
\Bigr]^{(2)}\cr
Z_{11}=&-\sum_k\sqrt{3}\hat k
\Bigl\lbrace\matrix{1&2&1\cr\ell_1&\ell_2&k}\Bigr\rbrace
\Bigl[Y^{\ell_1}(1)\otimes\bigl[Y^{\ell_2}(2)\nabla_2\bigr]^{(k)}
\Bigr]^{(2)}
}\eqno{8.8}$$
Combining this with with eq.(8.7), we can write the matrix element
contributions for the various form factors ($J=0$, $J=1$, and $J=2$) as
$$\eqalign{
<(1\bar 2)_{\lambda}\vert V^{R,LS2}_{(J=0)}&\vert (3\bar 4)_{\lambda}>=
-{\sqrt{5}\over{24}}{4\pi\over{2\lambda+1}}(-)^{(k_1+k_4)}{2\over{\pi}}
\int q^2dq \tilde V^{LS20}(q)
\sum_{\ell,\kappa}\hat\kappa
\Bigl\lbrace\matrix{1&1&2\cr\kappa&\ell&\lambda}\Bigr\rbrace\cr
\Bigl(
&<1\Vert \bigl[(Y^{\ell}\otimes\bigl[\vec\nabla\otimes\vec\nabla
\bigr]^{(2)})^{(\kappa)}\vec\sigma\bigr]^{(\lambda)}\Vert 2>
<4\Vert (Y^{\ell}\vec\sigma)^{(\lambda)}\Vert 3>\cr
+&<1\Vert (Y^{\ell}\vec\sigma)^{(\lambda)}\Vert 2>
<4\Vert \bigl[(Y^{\ell}\otimes\bigl[\vec\nabla\otimes\vec\nabla
\bigr]^{(2)})^{(\kappa)}\vec\sigma\bigr]^{(\lambda)}\Vert 3>\Bigr)}\eqno{8.9}$$
$$\eqalign{
<(1\bar 2)_{\lambda}\vert V^{R,LS2}_{(J=0)Z=2}&\vert (3\bar 4)_{\lambda}>=
-{5\over{12}}{4\pi\over{2\lambda+1}}(-)^{(k_1+k_4)}
\sum_{\ell,\ell_a,\ell_b}(-)^{(\ell_a+\ell_b)}
\hat \ell_a\hat\ell_b
\Bigl\lbrace\matrix{1&1&2\cr\ell_a&\ell_b&\ell}\Bigr\rbrace
\Bigl\lbrace\matrix{1&1&2\cr\ell_a&\ell_b&\lambda}\Bigr\rbrace\cr
&{2\over{\pi}}
\int q^2dq \tilde V^{LS20}(q)
\times <1\Vert\bigl[(Y^{\ell}\vec\nabla)^{(\ell_a)}\vec\sigma\bigr]
^{(\lambda)}\Vert 2>
<4\Vert\bigl[(Y^{\ell}\vec\nabla)^{(\ell_b)}\vec\sigma\bigr]
^{(\lambda)}\Vert 3>\Bigr)
}\eqno{8.10}$$
The contributions from terms $Z_{10}$, and $Z_{11}$, the $J=1$ terms
are
$$\eqalign{
<(1\bar 2)_{\lambda}\vert V^{R,LS2}_{(J=1)}\vert (3\bar 4)_{\lambda}>=
{5\over{\sqrt{12}}}{4\pi\over{2\lambda+1}}(-)^{(k_1+k_4)}
\sum_{\ell_a,\ell_b,\kappa}&(-)^{\kappa}\hat\kappa <\ell_a 0\ell_b 0|10>
\Bigl\lbrace\matrix{1&1&2\cr \ell_b&\kappa &\ell_a}\Bigr\rbrace
\Bigl\lbrace\matrix{1&1&2\cr \ell_b&\kappa &\lambda}\Bigr\rbrace\cr
{2\over{\pi}}\int q^2dq \bar V^{LS2,1}(q)\hat \ell_a
\hat \ell_b
(-)^{(\ell_a-\ell_b-1)/2}\Bigl(&
<1\Vert \bigl[(Y^{\ell_a}\vec\nabla)^{(\kappa)}\vec\sigma
\bigr]^{(\lambda)}\Vert 2><4\Vert 
(Y^{\ell_b}\vec\sigma)^{(\lambda)}\Vert 3>\cr
+<1\Vert& (Y^{\ell_b}\vec\sigma)^{(\lambda)}\Vert 2>
<4\Vert \bigl[(Y^{\ell_a}\vec\nabla)^{(\kappa)}\vec\sigma
\bigr]^{(\lambda)}\Vert 3>\Bigr) }\eqno{8.11}$$
Finally, the $J=2$ terms are from $Z_1$, $Z_2$, $Z_3$, $Z_7$, $Z_8$,
and $Z_9$.
$$\eqalign{
<(1\bar 2)_{\lambda}\vert V^{R,LS2}_{(J=2)}\vert (3\bar 4)_{\lambda}>=
{4\pi\over{2\lambda+1}}(-)&^{(k_1+k_4)}{2\over{\pi}}
\int q^2dq \tilde V^{LS2,2}(q)
\sum_{\ell_1,\ell_2}\hat \ell_1\hat\ell_2
<\ell_10\ell_20|20>(-)^{(\ell_1+\ell_2)/2}\cr
\biggl\lbrace\sum_{\kappa}{\sqrt{70}\over{24}}\hat\kappa
\Bigl\lbrace\matrix{1&1&2\cr\kappa&\ell_1&\lambda}\Bigr\rbrace&
\Bigl\lbrace\matrix{2&2&2\cr\ell_1&\ell_2&\kappa}\Bigr\rbrace
(-)^{(\kappa+\ell_1)}\cr
\Bigl(&
<1\Vert\Bigl[\bigl[Y^{\ell_2}\otimes\bigl[\vec\nabla\otimes
\vec\nabla\bigr]^{(2)}\bigr]^{(\kappa)}\vec\sigma\Bigr]^{(\lambda)}
\Vert 2><4\Vert(Y^{\ell_1}\vec\sigma)^{(\lambda)}\Vert 3>\cr
&+<1\Vert(Y^{\ell_1}\vec\sigma)^{(\lambda)}\Vert 2>
<4\Vert\Bigl[\bigl[Y^{\ell_2}\otimes\bigl[\vec\nabla\otimes
\vec\nabla\bigr]^{(2)}\bigr]^{(\kappa)}\vec\sigma\Bigr]^{(\lambda)}
\Vert 3>\Bigr)\cr
-{\sqrt{2}\over{24}}
\Bigl\lbrace\matrix{1&1&2\cr\ell_1&\ell_2&\lambda}\Bigr\rbrace\Bigr(&
<1\Vert\Bigl[\bigl[Y^{\ell_2}\otimes\bigl[\vec\nabla\otimes
\vec\nabla\bigr]^{(0)}\vec\sigma\Bigr]^{(\lambda)}\Vert 2>
<4\Vert (Y^{\ell_1}\vec\sigma)^{(\lambda)}\Vert 3>\cr
&+<1\Vert (Y^{\ell_1}\vec\sigma)^{(\lambda)}\Vert 2>
<4\Vert\Bigl[\bigl[Y^{\ell_2}\otimes\bigl[\vec\nabla\otimes
\vec\nabla\bigr]^{(0)}\vec\sigma\Bigr]^{(\lambda)}\Vert 3>\Bigr)
\biggr\rbrace}\eqno{8.12}$$
$$\eqalign{
<(1\bar 2)_{\lambda}\vert V^{R,LS2}_{(J=2)}\vert (3\bar 4)_{\lambda}>=&
-{4\pi\over{2\lambda+1}}(-)^{(k_1+k_4)}{2\over{\pi}}
\int q^2dq \tilde V^{LS2,2}(q)
\sum_{\ell_1,\ell_2}\hat \ell_1\hat\ell_2
<\ell_10\ell_20|20>(-)^{(\ell_1+\ell_2)/2}\cr
\sum_{\ell_a,\ell_b}
\hat \ell_a\hat\ell_b&
\Bigl\lbrace\matrix{1&1&2\cr\ell_a&\ell_b&\lambda}\Bigr\rbrace
\biggl( \sqrt{{2\over{3}}}{1\over{6}}
\Bigl\lbrace\matrix{\ell_2&\ell_1&2\cr\ell_a&\ell_b&\lambda}
\Bigr\rbrace
+\sqrt{14}{5\over{6}}
(-)^{(\ell_b+\ell_2)}
\Biggl\lbrace\matrix{\ell_1&\ell_2&2\cr 1&1&2\cr
\ell_a&\ell_b&2}\Biggr\rbrace\biggr)\cr
&<1\Vert\bigl[(Y^{\ell_1}\vec\nabla)^{(\ell_a)}\vec\sigma\bigr]
^{(\lambda)}\Vert 2>
<4\Vert\bigl[(Y^{\ell_2}\vec\nabla)^{(\ell_b)}\vec\sigma\bigr]
^{(\lambda)}\Vert 3>}
\eqno{8.13}$$
\vskip 0.25in

\noindent
{\bf 9. The isospin component}
\smallskip

\noindent
The isospin dependence of the matrix elements is best worked out using
the ($pp$) coupled matrix elements. In that case all we have to do is to
work out the expectation value of the isospin operator. We do this here
for the four isospin operators that appear in the Argonne $v_{18}$ potential 
of Wiringa {\it et al}~[1] and the four possible matrix elements 
(${\bf 1},\vec\tau_1\vec\tau_2,
T_{12}=3\tau_{z1}\tau_{z2}-\vec\tau_1\vec\tau_2,$ and
$\tau_{z1}+\tau_{z2}$):
$$\matrix{
<(pp)|{\bf 1}|(pp)>&=&+1 &~~,~~& <(pp)|\tau_{z1}+\tau_{z2}|(pp)>&=&+2\cr
<(pn)|{\bf 1}|(pn)>&=&+1 &~~,~~& <(pn)|\tau_{z1}+\tau_{z2}|(pn)>&=&~~0\cr
<(pn)|{\bf 1}|(np)>&=&~0 &~~,~~& <(pn)|\tau_{z1}+\tau_{z2}|(np)>&=&~~0\cr
<(nn)|{\bf 1}|(nn)>&=&+1 &~~,~~& <(nn)|\tau_{z1}+\tau_{z2}|(nn)>&=&-2}$$
$$\matrix{
<(pp)|\vec\tau_1\vec\tau_2|(pp)>&=&+1 &~~,~~&<(pp)|T_{12}|(pp)>&=&+2\cr
<(pn)|\vec\tau_1\vec\tau_2|(pn)>&=&-1 &~~,~~&<(pn)|T_{12}|(pn)>&=&-2\cr
<(pn)|\vec\tau_1\vec\tau_2|(np)>&=&+2 &~~,~~&<(pn)|T_{12}|(np)>&=&-2\cr
<(nn)|\vec\tau_1\vec\tau_2|(nn)>&=&+1 &~~,~~&<(nn)|T_{12}|(nn)>&=&+2}$$
Thus we calculate the matrix elements \hfill\break
for $<(pp)|V|(pp)>$
$$<(12)|V|(34)>=<(12)|V+V^{\tau\tau}+2V^{tT}+2V^{\tau z}+V^{Coul}
|(34)>$$
for $<(pn)|V|(pn)>$
$$<(12)|V|(34)>=<(12)|V-V^{\tau\tau}-2V^{tT}|(34)>$$
for $<(pn)|V|(np)>$
$$<(12)|V|(34)>=<(12)|2V^{\tau\tau}-2V^{tT}|(34)>$$
and for $<(nn)|V|(nn)>$
$$<(12)|V|(34)>=<(12)|V+V^{\tau\tau}+2V^{tT}-2V^{\tau z}|(34)>$$

\beginsection{10. CM-correction terms}

From the center-of-mass (CM) corrections we have two additional terms that
must be included in the two-body matrix elements of the interaction.
These are:
$$-{{\vec p_1\cdot\vec p_2}\over{mA}}=+{{\hbar^2\vec\nabla_1\cdot\vec\nabla_2}
\over{mA}}$$
and
$${{\vec r_1\cdot\vec r_2} \over{mA}}$$
These can be computed using eq.(2.1). These matrix elements contribute only
for $\lambda=1$ and can be computed with eq.(3.3) resulting in
$$-{{1}\over{mA}}<1\bar 2|\vec p_1\vec \cdot p_2|3\bar 4>={\hbar^2\over{mA}}
(-)^{k_1}{1\over{\sqrt{3}}}<1\Vert \nabla\Vert 2>
(-)^{k_4}{1\over{\sqrt{3}}}<4\Vert \nabla\Vert 3>$$
and
$${{1}\over{mA}}<1\bar 2|\vec r_1\vec \cdot r_2|3\bar 4>={{1}\over{mA}}
(-)^{k_1}{1\over{\sqrt{3}}}<1\Vert \vec r \Vert 2>
(-)^{k_4}{1\over{\sqrt{3}}}<4\Vert \vec r \Vert 3>$$
Here
$$(-)^{k_1}{1\over{\sqrt{3}}}<1\Vert\vec\nabla\Vert 2>=
{{\hat j_1\hat j_2}\over{3}} <j_1{1\over{2}}j_2{-1\over{2}}|10>
<R_1(r)|{d\over{dr}}+{1\over{2}}(\ell_2-\ell_1)(3\ell_2-\ell_1+1
) {1\over{r}}|R_2(r)>$$
and
$$(-)^{k_1}{1\over{\sqrt{3}}}<1\Vert \vec r\Vert 2>=-
{{\hat j_1\hat j_2}\over{3}} <j_1{1\over{2}}j_2{-1\over{2}}|10>
<R_1(r)|r|R_2(r)>$$
\vskip 0.25in

\noindent
{\bf 11. The radial integrals}
\smallskip

\noindent
We assume that each wave function is expanded in harmonic oscillator
functions corresponding to the oscillator length parameter $ok$.
$$R_{i,l}(r)=\sum_n A^i_nH_{n,l}(r)\eqno{10.1}$$
As such each of the radial wave functions can be written as a
polynomial multiplied by a gaussian of argument $-(\sqrt{2}r/ok)^2$.
We now introduce the variable $x=\sqrt{2}r/ok$. With this, each
radial wave function can be written in terms of $x$ as
$$R_{i,l}(x)=\bar R_{i,l}(x)e^{-x^2/4}\eqno{10.2}$$
where $\bar R_{i,l}$ is a ploynomial in $x$. Generally, we define the
barred functions as those without the exponential. In terms of $x$ the wave
functions are normalized such that
$$\int R_{i,l}^2(x)x^2dx=\Biggl({\sqrt{2}\over{ok}}\Biggr)^3\eqno{10.3}$$
In order to carry out the Fourier transform we use the fact that
the Fourier transforms of the harmonic oscillator functions are
again harmonic oscillator functions in $k$-space. Thus we expand the
product of two radial functions as
$$R_{i,l_i}(x)R_{j,l_j}(x)=\sum_n B^{i,j}_n H_{n,L}(x)\eqno{10.4}$$
Using Gaussian integration, the expansion coefficients can be written
as 
$$B^{i,j}_n=\sum_k \bar H_{n,L}(x_k)\bar
R_{i,l_i}(x_k)\bar R_{j,l_j}(x_k)x_k^2w_k\eqno{10.5}$$
with this, we write the integral
$$\int R_{i,l_i}(r)j_L(qr)R_{j,l_j}(r)r^2dr=\sum_n
 B^{i,j}_nH_{n,L}(\bar q)\eqno{10.6}$$
 where $\bar q=\sqrt{2}ok~q$.
 The radial integrals can now be computed according to
 $$\eqalign{{2\over{\pi}}
 \int q^2dq ~V(q)& \langle R_{i,l_i}\vert j_L(qr)\vert R_{j,l_j}\rangle
                    \langle R_{s,l_s}\vert j_K(qr)\vert R_{t,l_t}\rangle=\cr
 & =\sum_n\sum_mB^{i,j}_nB^{s,t}_m{2\over{\pi}}\int q^2dq~V(q)H_{n,L}(\bar q)
  H_{m,K}(\bar q)\cr
 & =\colon \sum_n\sum_mB^{i,j}_nB^{s,t}_m I^{L,K}_{n,m}}\eqno{10.7}$$
 Inserting the expression for the expansion coefficients $B$ we find
$$\eqalign{ {2\over{\pi}}
 \int q^2dq ~V(q)& \langle R_{i,l_i}\vert j_L(qr)\vert R_{j,l_j}\rangle
                    \langle R_{s,l_s}\vert j_K(qr)\vert R_{t,l_t}\rangle=\cr
 & ={2\over{\pi}}\sum_k\sum_q \bar
R_{i,l_i}(x_k)\bar R_{j,l_j}(x_k)x_k^2w_k \bar
R_{s,l_s}(x_q)\bar R_{t,l_t}(x_q)x_q^2w_q
\sum_{m,n} I^{L,K}_{n,m}\bar H_{n,L}(x_k)\bar H_{m,K}(x_q)\cr
&=\colon~\sum_k\sum_q \bar
R_{i,l_i}(x_k)\bar R_{j,l_j}(x_k)\bar
R_{s,l_s}(x_q)\bar R_{t,l_t}(x_q) ~S^{L,K}_{k,q}}\eqno{10.8}$$
With this notation all terms can be computed as
$$\eqalign{
<(1\bar 2)_{\lambda}|V^{tot}|(3\bar 4)_{\lambda}>=\sum_{i,j}&
\bar R_1(x_i)\bar R_2(x_i)S^1_{i,j}\bar R_3(x_j)\bar R_4(x_j)\cr
+&\bar R_1(x_i)\bar R_2'(x_i)S^2_{i,j}\bar R_3(x_j)\bar R_4(x_j)\cr
+&\bar R_1(x_i)\bar R_2(x_i)S^3_{i,j}\bar R_3'(x_j)\bar R_4(x_j)\cr
+&\bar R_1(x_i)\bar R_2''(x_i)S^4_{i,j}\bar R_3(x_j)\bar R_4(x_j)\cr
+&\bar R_1(x_i)\bar R_2(x_i)S^5_{i,j}\bar R_3''(x_j)\bar R_4(x_j)\cr
+&\bar R_1(x_i)\bar R_2'(x_i)S^6_{i,j}\bar R_3'(x_j)\bar R_4(x_j)}\eqno{10.9}$$
Correspondingly, the exchange contribution can be written as
$$\eqalign{
<(1\bar 2)_{\lambda}|V^{tot}|(3\bar 4)_{\lambda}>=\sum_{i,j}&
\bar R_1(x_i)\bar R_3(x_i)E^1_{i,j}\bar R_2(x_j)\bar R_4(x_j)\cr
+&\bar R_1(x_i)\bar R_3'(x_i)E^2_{i,j}\bar R_2(x_j)\bar R_4(x_j)\cr
+&\bar R_1(x_i)\bar R_3(x_i)E^3_{i,j}\bar R_2'(x_j)\bar R_4(x_j)\cr
+&\bar R_1(x_i)\bar R_3''(x_i)E^4_{i,j}\bar R_2(x_j)\bar R_4(x_j)\cr
+&\bar R_1(x_i)\bar R_3(x_i)E^5_{i,j}\bar R_2''(x_j)\bar R_4(x_j)\cr
+&\bar R_1(x_i)\bar R_3'(x_i)E^6_{i,j}\bar R_2'(x_j)\bar R_4(x_j)}\eqno{10.10}$$
As the marices $S$ and $E$ only depend on the angular momenta 
involved but not on the quantum numbers $n_1$,$n_2$,$n_3$, and
$n_4$, the strategy is to compute the matrices first and then
compute all the matrix elements with the same angular momentum 
combinations. This way a large number of matrix elements can be computed
in which orbits differ only by the quantum number $n$.

\vskip 0.2in
\noindent
Note. This strategy is essential in computing the G-matrix interaction in
calculating the term:
$$\sum_K \sum_{p_3,p_4}Z^K_{p_3p_4,h_1h_2}<(p_3p_4)_K|V|(p_1p_2)_K>$$

\beginsection{References}

\frenchspacing
\item{[1]}{R.B.Wiringa, V.G.J.Stoks, and R.Schiavilla, 
     {{\it Phys. Rev.} C~{\bf 51}, 38 \rm (1995)}}
\item{[2]}{S.Krewald, V.Klemt, J.Speth, and A.Faessler, 
     {\it On the Use of Skyrme-Forces in Selfconsistent RPA-Calculations}, 
     J\"ulich, Germany}
\item{[3]}{A.R. Edmonds, 
     {\it Angular Momentum in Quantum Mechanics}, 
     Princeton University Press, Princeton, 3rd ed., 1965}

\vfill
\eject
\bye